\documentclass[10pt, conference, a4paper]{IEEEtran}

\usepackage{url}
\usepackage{graphicx}
\usepackage{color}
\usepackage{subcaption}
\usepackage{hyperref}
\usepackage[T1]{fontenc}
\usepackage{longtable}
\usepackage{tabularx}
\usepackage{array}
\usepackage{cleveref}
\usepackage{multirow}
\usepackage{color}
\usepackage{enumitem}
\usepackage[table]{xcolor}
\usepackage[
  pass,
]{geometry}

\definecolor{applegreen}{rgb}{0.55, 0.71, 0.0}

\title{Where is My Next Hop ? \\ The Case of Indian Ocean Islands}

\author{
\IEEEauthorblockN{Xavier Nicolay\IEEEauthorrefmark{1},
                  Rehan Noordally\IEEEauthorrefmark{1},
                  Pascal Anelli\IEEEauthorrefmark{1},
                  Nour Mohammad Murad\IEEEauthorrefmark{2},
                  and
                  Tahiry Razafindralambo\IEEEauthorrefmark{1}
                 } \\

\IEEEauthorblockA{\IEEEauthorrefmark{1}Laboratoire d'Informatique et de Math\'ematiques}
\IEEEauthorblockA{\IEEEauthorrefmark{2}Laboratoire d'Energ\'etique, d'Electronique et Proc\'ed\'es}
\IEEEauthorblockA{University of Reunion Island\\
15 Avenue Ren\'e Cassin\\
97490 Sainte Clotilde\\
France\\email: \{xavier.nicolay, rehan.noordally, pascal.anelli, nour.murad, tahiry.razafindralambo\}@univ-reunion.fr}}

\begin{document}
\maketitle

\begin{abstract}

Internet has become a foundation of our modern society. However, all regions or countries do not have the same Internet access regarding quality especially in the Indian Ocean Area (IOA). To improve this quality it is important to have a deep knowledge of the Internet physical and logical topology and associated performance. However, these knowledges are not shared by Internet service providers. In this paper, we describe a large scale measurement study in which we deploy probes in different IOA countries, we generate network traces, develop a tool to extract useful information and analyze these information. We show that most of the IOA traffic exits through one point even if there exists multiple exit points.

\end{abstract}




\section{Introduction}
\label{sec:introduction}

Internet is fundamental to our society as it provides important services ranging from safety and security services to entertainment. Internet was designed to carry applications data with no time constraint and limited user interactions. Nowadays, several applications focus on users' interactions and timely content delivery is critical. As the Internet connectivity is expected to improve, the user experience is also expected to improve. However, all countries or regions are not equal from the Internet access and performance  point of view.


The Internet end-to-end connectivity is not improving, as demonstrated by Lee~{\it et al.} in~\cite{Lee2010} and Cardozo {\it et al.} in~\cite{Cardozo2012}. Within the last decade, changes in the Internet path and bufferbloat issues have worsen the TCP's congestion control~\cite{Gettys2011}. However, beyond bandwidth, low latency is required for new Internet applications.
Some geographic area such as The Indian Ocean Area (IOA) including Madagascar, Mauritius, Mayotte, Reunion Island and Seychelles have poorly meshed Internet topology and low performance. In \cite{Noordally2016, Noordally2017}, Noordally~{\it et al.} focus on Reunion Island Internet connectivity and performance. In this paper, we focus on the whole Indian Ocean Area (IOA) region.


Improving the Internet access for the IOA is important since it can break the Internet silo in this region and help the development of these countries. Briscoe {\it et al.} wrote a survey in \cite{latency2014} that describes the factors of latency and proposes some solutions to reduce it. It includes exploiting path diversity to select the shortest path and load-balancing to prevent congestion. To evaluate the possible implementation of these solutions, we need a deep understanding of the physical and logical Internet topology in this area.  

To the best of our knowledge, no such large scale study has been carried out by the scientific community in the IOA region. This paper tries to fill this gap by:
\begin{itemize}
\item Deploying 16 probes in  different countries located in IOA: Madagascar (MG), Mauritius (MU), Reunion Island (RE), Seychelles (SC) and Mayotte (YT).
\item Generating $4,480,000$ \emph{traceroute} traces using randomly selected IPv4 addresses during a one month measurement campaign.
\item Developing a tool to extract the logical topology of the Internet in the IOA region based on IP localization and ICMP variant of Paris-Traceroute.
\item Analyzing the traces and provide some insight on issues of Internet Access in the IOA regarding exit point of each country, path length and geographical distances.
\end{itemize}

Our main contribution is to analyze and understand the traffic information from the IOA islands to identify the bottleneck of the Internet traffic in this region.

The remainder of this paper is organized as follows. Section~\ref{sec:description} describes the topology of the submarine cables connecting Indian Ocean's Island to the Internet as well at the {\em Internet eXchange Point} (IXP).
The results are analyzed in section~\ref{sec:results}.  Section~\ref{sec:related} reviews the related work. Finally, we conclude in section~\ref{sec:conclusion}.



    
    

\section{Background}\label{sec:description}

\begin{figure}[ht!]
\centering
\includegraphics[width=0.48\textwidth]{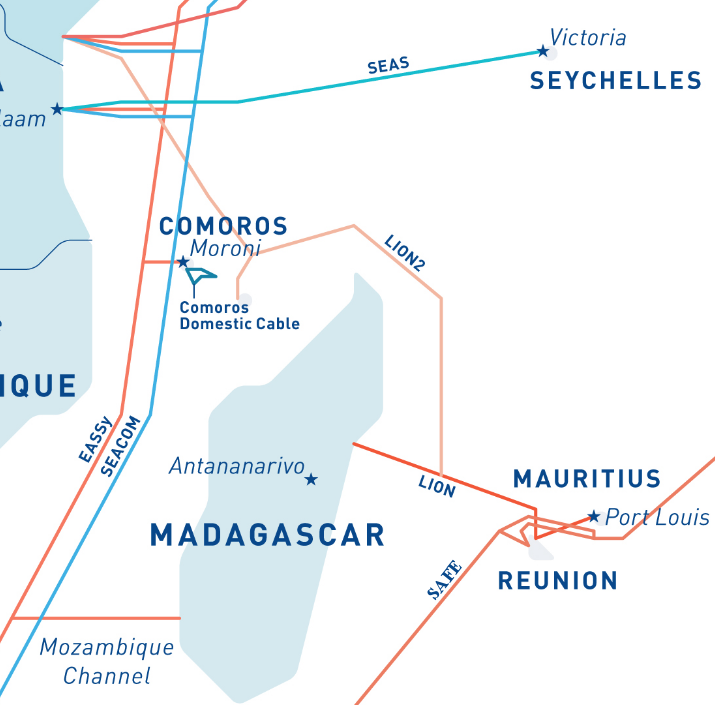}
\caption{Mascarene islands submarine cable. \cite{cablesmap}}
\label{map:cable}
\end{figure}

The map in figure~\ref{map:cable}\footnote{Source: www.submarinecablemap.com} shows that each Island is connected to the Internet with one or more submarine cables. We can notice that LION/LION2 cable provides a link between Mayotte, Madagascar, Reunion and Mauritius. These 4 islands have also an IXP. This equipment can be used by each~\emph{Internet Service Provider (ISP)} connected to it to exchange their traffic. 

We know that:
   \begin{itemize}
      \item{ Each Country / Island is connected to one or more Submarines cables (see figure~\ref{map:cable}), so we know the real topology.}
      \item{ There is 4 IXP (Mayotte, Reunion, Mauritius, Madagascar). See AXIS Project~\cite{axis} for more information.}
      \item{ There are a few of ISP, none are present everywhere: ComoresTelecom, Emtel, CEB FiberNET Co Ltd, Blueline, Telma, Canal + Telecom, Orange, SRR (SFR R\'eunion), STOI, Telco OI (Only), Zeop, Airtel, Cable \& Wireless, Intelvision and Kokonet.}
   \end{itemize}
   
We do not know:
   \begin{itemize}
        \item {The exact interconnection between the IXPs. This information is very useful to evaluate logical topology of the network.}
        \item {The logical path of a TCP/IP session. This information is related to the core objective of our paper. We aim at analyzing the Internet access of IOA islands.}
        \item {The regional traffic in percent of international traffic. This information could help us in our analysis. Indeed, Internet access performance is strongly correlated with traffic shapes.}
        \item {The capacity of each Internet Service Provider. This information could provide us some intuition on the traffic and peering policy of each ISP.}
	\end{itemize}

In this paper, based on our knowledge of the IOA islands Internet architecture, we aim at analyzing and understanding the traffic information from the IOA islands to identify the bottleneck of the Internet traffic in this region.  

\section{Measurement operations}
\label{sec:methodo}
We study the Internet connectivity of Islands localized in the Indian Ocean Area according to the delay and the network paths. To do so we collected \emph{traceroute} traces between some islands of Indian Ocean and $10,000$ destinations distributed worldwide. \\

Our active measurements made from Indian Ocean involve $16$ raspberry pi~\cite{raspberrypi} probes distributed over the 5 countries: 
2 hosted at Madagascar, 1 at Mauritius, 1 at Mayotte, 11 at Reunion Island, and 1 at Seychelles. Our trace includes measurements performed from March $22^{nd}$ 2017 to April $22^{nd}$ 2017. We created a random set of $1,000,000$ public IPv4 addresses among which only $83,850$ responded to ICMP Echo request.

This new set was geo-referenced by country. The second column of table~\ref{tab:ip_repartition} shows the geographical distribution of these IPv4 addresses and the third column shows the actual distribution of the IPv4 addresses provided by the website  \url{https://www.countryipblocks.net}\footnote{The distribution was retrieved from the \emph{countryipblocks} website on the $4^{th}$ of May 2016}. The two distributions are distant from one another. To respect the actual representation, we have decided to use the second one.
Among these $83,850$ IPv4 addresses, we selected a subset of $10,000$ addresses that fits the actual geographical distribution. 
\begin{table}[ht!]
  \centering
  \caption{Geographical distribution of our $83,850$ randomly obtained IPv4 addresses and the actual geographical distribution provided by the \emph{CountryIpBlocks} website.}
  \begin{tabular}{|p{0.15\textwidth}|p{0.08\textwidth}|p{0.17\textwidth}|}
    \hline
    \multicolumn{1}{|c|}{Continent}&\multicolumn{1}{c|}{Random}&\multicolumn{1}{c|}{CountryIPBlocks}\\
    \hline
    \multicolumn{1}{|c|}{Africa (AF)} & \multicolumn{1}{c|}{0.95\%} & \multicolumn{1}{c|}{2.59\%} \\
    \hline
    \multicolumn{1}{|c|}{Asia (AS)} & \multicolumn{1}{c|}{32.96\%} & \multicolumn{1}{c|}{23.34\%} \\
    \hline
    \multicolumn{1}{|c|}{Europe (EU)} & \multicolumn{1}{c|}{28.99\%} & \multicolumn{1}{c|}{20.7\%} \\
    \hline
    \multicolumn{1}{|c|}{North America (NA)} & \multicolumn{1}{c|}{8.89\%} & \multicolumn{1}{c|}{47.55\%} \\
    \hline
    \multicolumn{1}{|c|}{Oceanie (OC)} & \multicolumn{1}{c|}{0.7\%} & \multicolumn{1}{c|}{1.55\%} \\
    \hline
    \multicolumn{1}{|c|}{South America (SA)} & \multicolumn{1}{c|}{6.30\%} & \multicolumn{1}{c|}{4.27\%} \\
    \hline
    \multicolumn{1}{|c|}{Other (bogons)} & \multicolumn{1}{c|}{21.19\%} & \multicolumn{1}{c|}{0.0\%} \\
    \hline
  \end{tabular}
  \label{tab:ip_repartition}
\end{table}

Each of our local probes was configured to perform a \emph{traceroute} toward all of the IP of our data set within one day. 
A probe started a new measurement every $8.64$s which lasted for an average of $28$s. The number of \emph{traceroutes} running simultaneously has been limited to $4$, resulting in a maximum bit-rate of $5,06Kb/s$, which is negligible compared to the available bandwidth which is at least of $128.33 Kb/s$ in Reunion Island \cite{Vergoz2013}. To further prevent the congestion induced by our measurements on the destination, the sequence of destinations to visit was randomized on each probe.
Our final data set contains a total of $4,480,000$ \emph{traceroute} traces.

\section{Tools involved}
\label{sec:tools}
\subsection{Traceroute tool}
The original \emph{traceroute}~\cite{rfc1393} developed by Malkin is known to produce inconsistent results in the context of load-balancing.  

To circumvent this issue, \emph{Paris-Traceroute} was created by the authors of~\cite{augustin2006}. Thus, TCP packets are sent instead of ICMP. In~\cite{Wenwei2006}, the authors compared the ICMP and TCP techniques. While they found that in most of the cases the results are similar, when the ratio between the mean \textit{Round-Trip Time} (RTT) and the minimum RTT tends to be large (beyond 20), the results of the TCP variant tend to be less stable. For this reason, we use the ICMP version of \emph{Paris-Traceroute} protocol in our experiment.

\subsection{Geolocation tool}
The coordinate of the IPv4 address were obtained with the database of \emph{RIPE NCC}~\cite{RIPENCC}. We used their API~\cite{API-RIPE} to retrieve information such country, latitude, longitude and AS about each the $83,850$ IPv4 addresses and each of the routers found during the \emph{traceroute} measurements.

In order to update the localization and enhanced the performance, our own MySQL database is used. We have two main tables: one with \{IP, Latitude, Longitude, Countries\} and one other with \{IP, mapx, mapy\}. 

We found out that some of the IPs were not properly geo-localized. We inferred an approximate geo-localization of the node according to the minimum delay from several probes distributed worldwide. Then, an IP was considered to be part of the same continent as the probes with which it had the closest delay. The database used for the geolocalization are used in our tool.

\subsection{Our original tool: \emph{Rtraceroute}}
As we want to handle more than 1 million traces, we develop our tool \cite{Nicolay2017-1} with C and threadpool \cite{Heber2013}. Maximum of parallelization are implemented and now the tool can parse about 4.5 millions in about 1 hour on a computer with 8 Cores (which is a part of an IBM 3650 on VMWare).\\
All traces are read and IP are geolocated. When the country change between 2 IP address, a link is created.
Bogon IP ( * or private IP) change nothing.
The tool generate two maps from an empty map of the world \cite{emptyWorld}.
The first one draws all links to a new map.
The second map is also created: one point of the link is the country studied (for example: Mauritius 
which has its coordinates \{x,y\} = \{2611,1569\} for the map and stored in the MySQL database).
Filtering the links with an extremity permit to show the next country-hop of a country.
In other words, we can simply discover the real connectivity of a country.

Our tool is available at \href{http://lim.univ-reunion.fr/rtraceroute}{http://t.univ-reunion.fr/414}

\subsection{Data filtering tool}
Our measurement campaign lead to a raw data set of $4,480,000$ \emph{traceroute} traces. But some traces are useless and need to be sanitized. We removed traces that met one of the following criteria: 
\begin{itemize}
\item the destination has not been reached
\item there is 3 following stars at the end of the trace
\item the presence of '!N' (network unreachable) or '!H' (host unreachable) marks due to \emph{Paris-Traceroute}
\item some corrupted trace (exception probe, empty trace, ...)
\item the loops (more than 200 hops)
\item the presence of IP whose countries are not present in the \emph{RIPE NCC} database This criterium was only applied for the geographic path analysis performed in section~\ref{sec:pathanal}
\end{itemize}
We obtain a new data set after filtering of $1,053,894$ clean \emph{traceroute} traces. Our Traces are available at \href{http://lim.univ-reunion.fr/rtraceroute/Data/data_GIIS2017.tar.gz}{http://t.univ-reunion.fr/411}.

\section{Results}\label{sec:results}

\subsection{Path length, geographical distance and Round Trip Time}\label{5.1}

\begin{figure}[ht!]
		\centering
    	\includegraphics[width=0.87\linewidth]{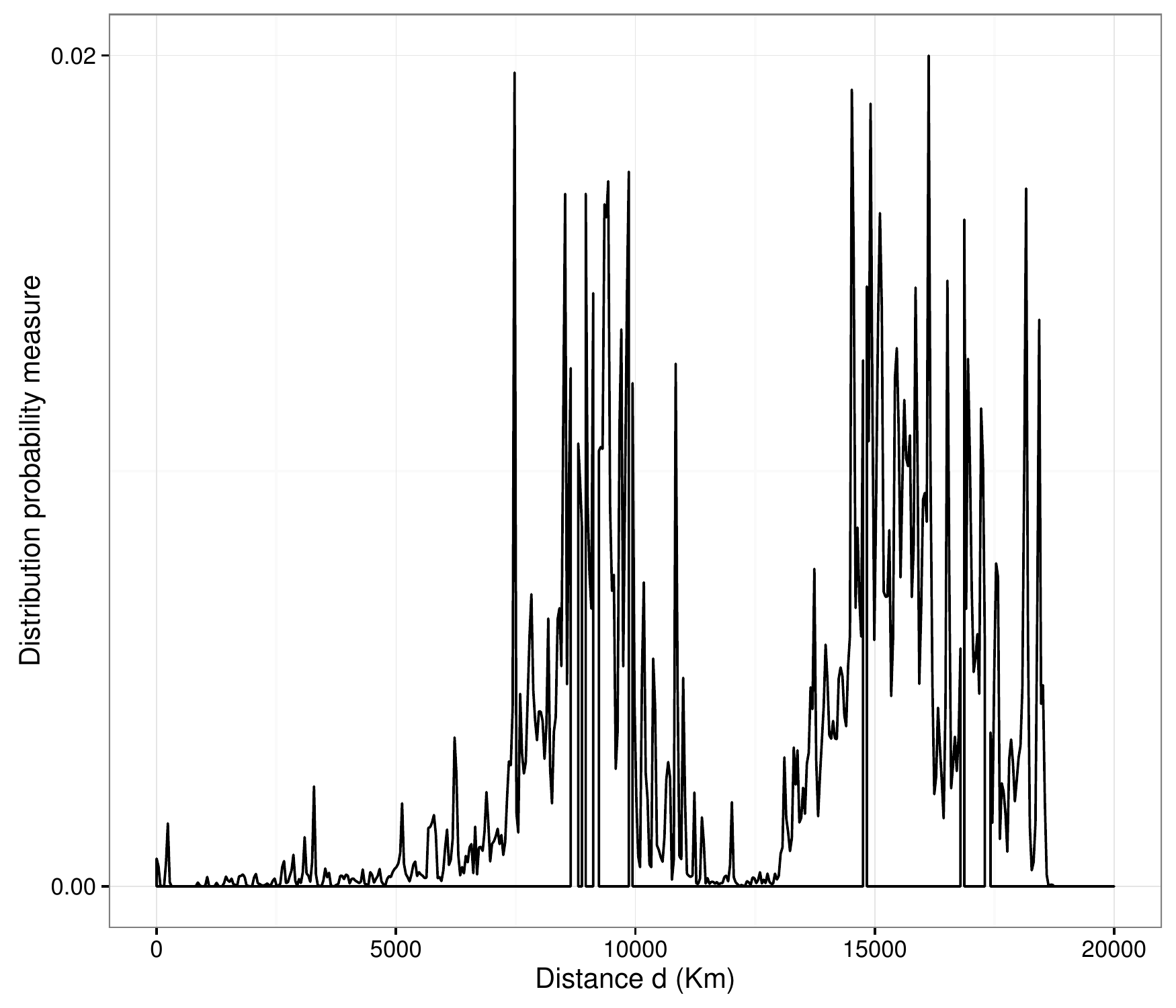}
    	\caption{Distribution of the distance between IOA's sources and world's destinations.} 
    	\label{fig:distance_zoi}
\end{figure}

The first set of results we obtained from our large scale experiments with the nature of traffic from the IOA islands to the rest of the world in figure~\ref{fig:distance_zoi}. This figure shows the distribution of the distance between IOA islands and their random destination in kilometers based IP location. The distribution result confirms the experiments settings given in table~\ref{tab:ip_repartition}. The two pics in the figure at 10,000 km and 15,000 km represents Europe plus Asia and North America respectively. 

\begin{figure}[ht!]
	\centering
	\includegraphics[width=0.70\linewidth, angle=-90]{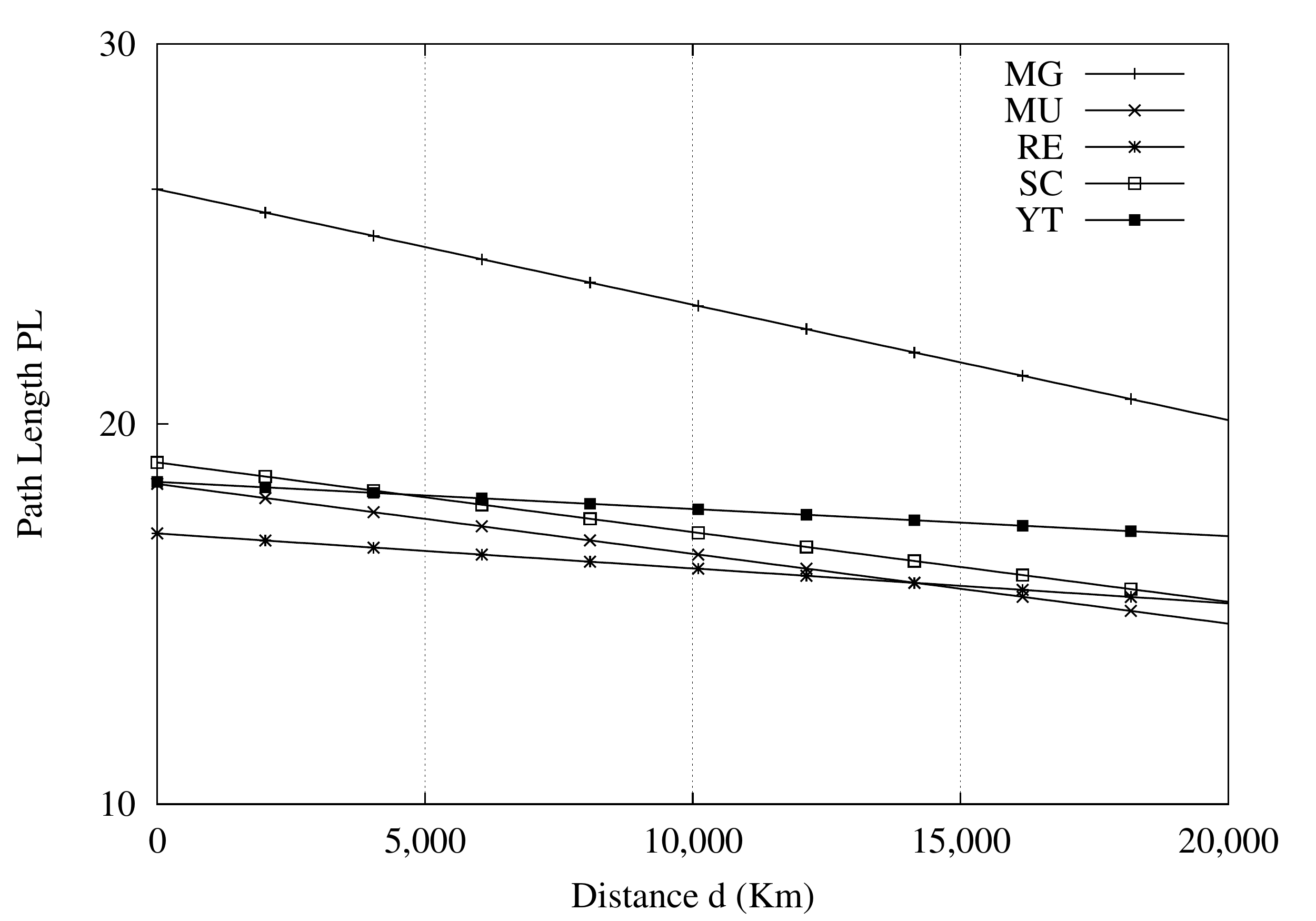}
	\caption{Dependence of the path length on geographical location.} 
	\label{fig:pl_d}
\end{figure}

In the second set of results, we plot the average path length in numbers of hops depending of the geographical distance, see figure~\ref{fig:pl_d}. The measure shows that the number of hops is stable depending relatively to the distance except for Madagascar.  It is important to notice here that on average even if the distance between the source IP address (an IOA island) is geographically close to the destination IP address (between 0 and 5,000 km) the number of hops remains high on average. 

A first interpretation of figure~\ref{fig:pl_d} is that the number of hops between an IOA island and another IOA island is on average the same as number of hops between an IOA island and an European country. This is even worse in the case of Madagascar where reaching an IOA island needs on average more hops than reaching an European country.\\


    
\begin{figure}[ht!]
	\centering
	\includegraphics[width=0.7\linewidth, angle=-90]{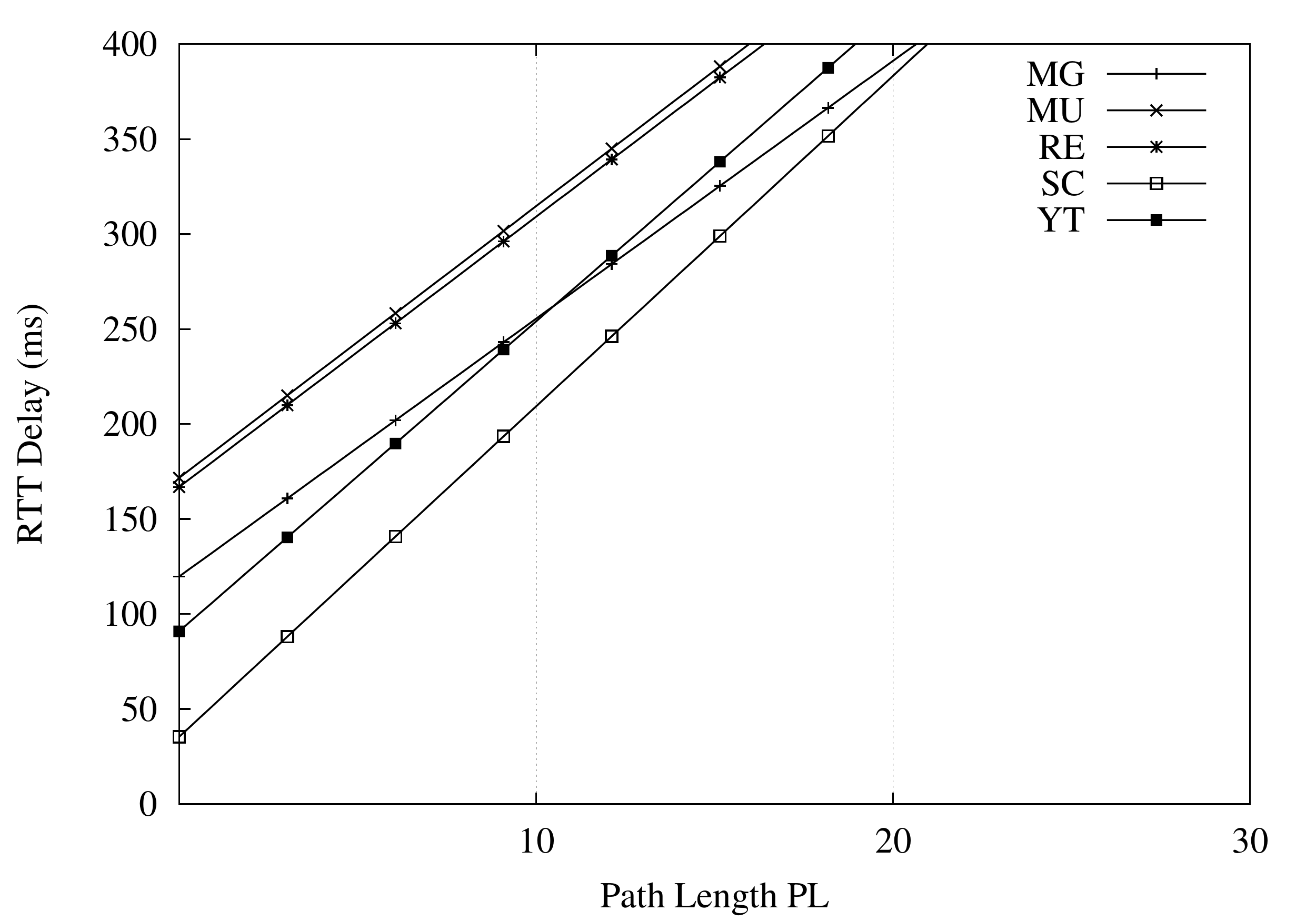}
	\caption{Dependence of the RTT and the path length.} 
	\label{fig:pl_rtt}
\end{figure}
\begin{figure}[ht!]
	\centering
	\includegraphics[width=0.67\linewidth, angle=-90]{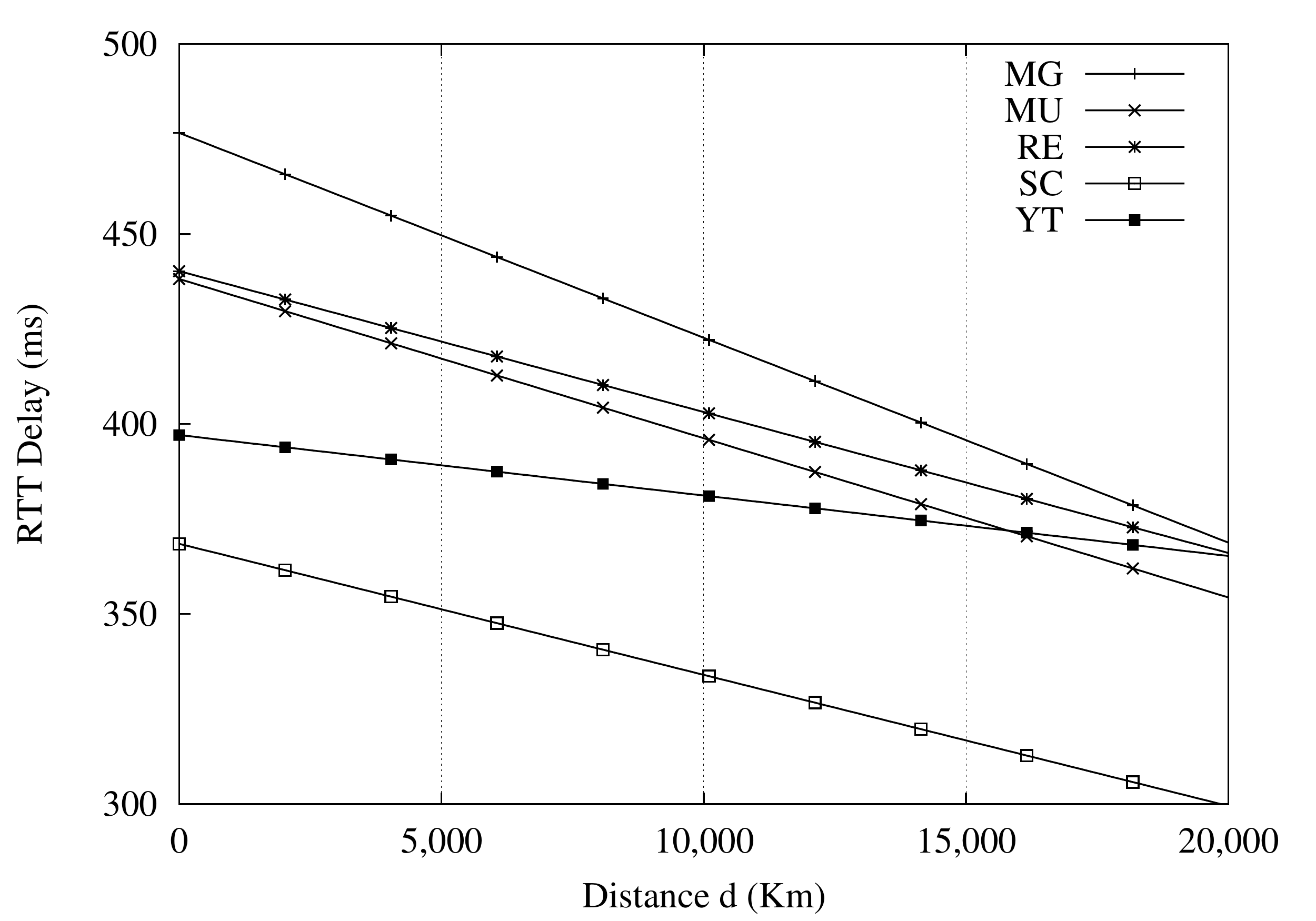}
	\caption{Dependence of the RTT on geographical location.} 
	\label{fig:d_rtt}
\end{figure}

Figure~\ref{fig:pl_rtt} plots the RTT depending on the number of hops. As expected, the RTT increases with the number of hops. However as shown in Figure~\ref{fig:d_rtt} the RTT is stable depending on the distance. This result shows that from an IOA island point of view, geographically close (small distance in kilometers) are harder to reach in terms of RTT and number of hops.  

The results in this subsection tend to show that geographical distances are not related to hop distances from the IOA islands point of view. They indicate that on average when traffic exits from an IOA island wherever the destination is (in terms of geographical position), this destination will be reached in a constant number of hops. 

\subsection{Path Analysis} \label{sec:pathanal}

\begin{figure}[ht!]
	\centering
	\includegraphics[height=0.47\textwidth, angle=-90]{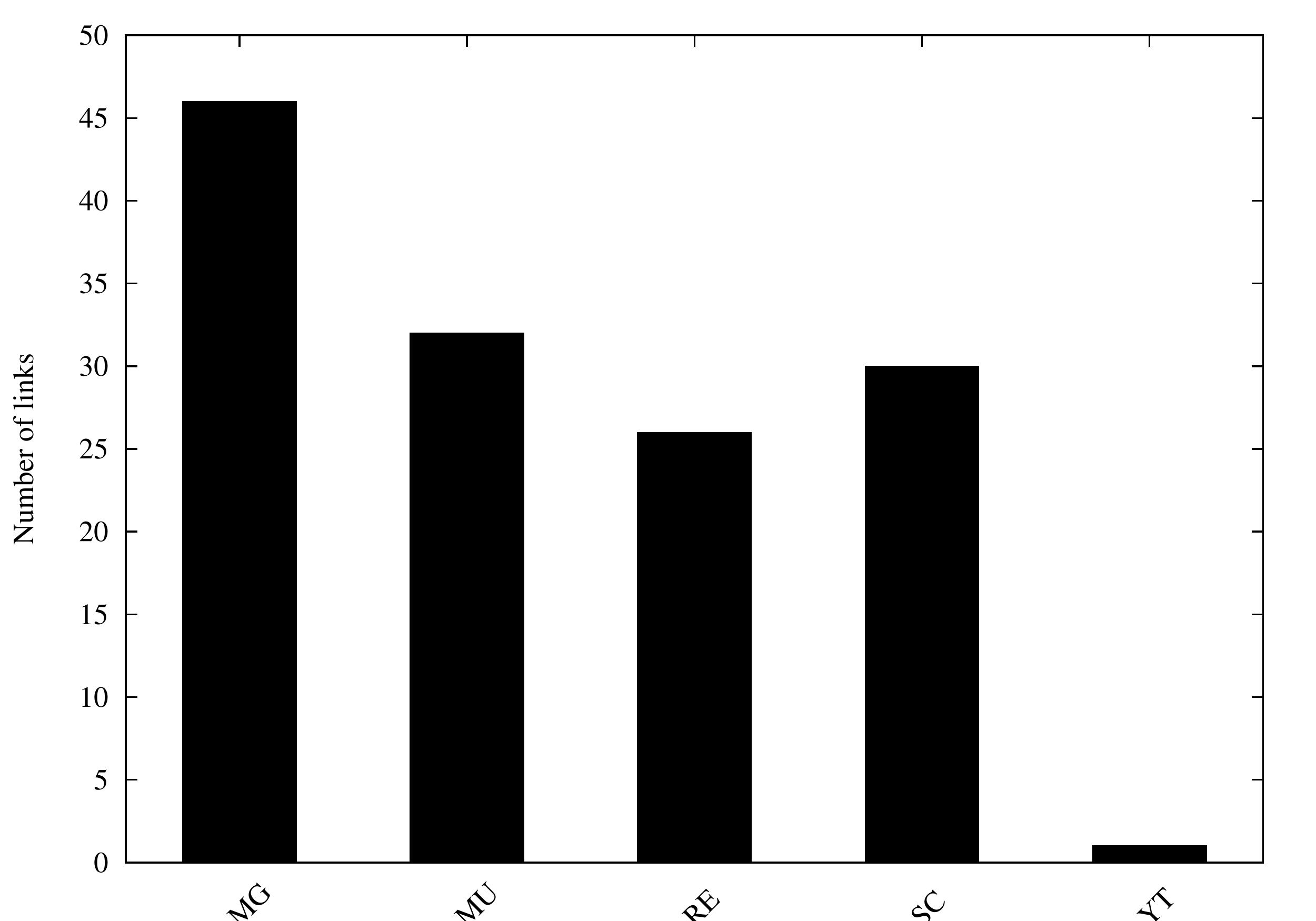}
	\caption{Number of Internet output links from each countries from IOA}
	\label{fig:output}
\end{figure}

In Figure~\ref{fig:output}, we plot the number of exit points for each IOA island. By exit point, we mean the first hop for each island outside its own country. This Figure shows that except for Mayotte which have only one exit point, each island have more than 20 exit points. This figure seems to indicate some path diversity for the Internet traffic of each island. In table~\ref{tab:exitpoint} the details of the exit points of each country are given. This table gives the ISO-3166 country code\footnote{\href{https://fr.wikipedia.org/wiki/ISO\_3166}{https://fr.wikipedia.org/wiki/ISO\_3166}} for the exit point and the percentage of traffic using this exit point. This table confirms the diversity of the exit points, except for Mayotte, but also shows that for most islands in the IOA more than 90\% of our data exits through only three countries (written in bold in the tabular).
Table~\ref{tab:exitpoint} also shows some asymmetrical behavior in routing and peering policies. For example, Mauritius (MU) appears in Madagascar's (MG) exit point but Madagascar does not appear in Mauritius'.    

\begin{table}[h]
\centering
\caption{Table of exit point from IOA}
\begin{center}
\begin{scriptsize}
\begin{tabular}{ | l | l | l | l | l |} 
\hline
\multicolumn{1}{|c|}{MG} & \multicolumn{1}{c|}{MU} & \multicolumn{1}{c|}{RE} & \multicolumn{1}{c|}{SC} & \multicolumn{1}{c|}{YT} \\ \hline
AE 0.0005\% & AZ 0.001\%   & AE 0.0001\%   & EU 0.0001\%  & \textbf{FR 100\%} \\
AU 0.0005\% & HU 0.001\%   & DK 0.0001\%   & KE 0.0001\%  & \\
FI 0.0005\% & IN 0.001\%   & TH 0.0001\%   & AZ 0.002\%   & \\
JP 0.0005\% & JP 0.001\%   & MQ 0.0005\%   & DZ 0.002\%   & \\
MU 0.0005\% & TW 0.001\%   & RO 0.0005\%   & ID 0.002\%   & \\
SA 0.0005\% & CA 0.003\%   & AR 0.0007\%   & IL 0.002\%   & \\
TR 0.0005\% & AE 0.004\%   & TW 0.0007\%   & NA 0.002\%   & \\
AT 0.001\%  & CO 0.004\%   & RS 0.0014\%   & SA 0.002\%   & \\
CH 0.001\%  & CZ 0.004\%   & CR 0.0017\%   & CO 0.01\%    & \\
GR 0.001\%  & RS 0.004\%   & BG 0.0019\%   & PK 0.011\%   & \\
NG 0.001\%  & SE 0.004\%   & SE 0.0019\%   & HK 0.013\%   & \\
HU 0.0015\% & TH 0.004\%   & AT 0.0020\%   & RO 0.018\%   & \\
IS 0.0015\% & BG 0.006\%   & HK 0.0030\%   & IT 0.023\%   & \\
MD 0.0015\% & GR 0.006\%   & JP 0.0049\%   & TH 0.034\%   & \\
PH 0.0015\% & PL 0.007\%   & SG  0.0073\%  & ZA 0.037\%   & \\
ES 0.002\%  & US 0.007\%   & NL 0.0103\%   & SE 0.038\%   & \\
KE 0.0025\% & ZM 0.007\%   & ZA 0.0117\%   & VE 0.039\%   & \\
IE 0.003\%  & RU 0.012\%   & IE 0.0121\%   & IE 0.040\%   & \\
IL 0.004\%  & UG 0.013\%   & RU 0.0241\%   & SG  0.043\%  & \\
AZ 0.0045\% & CR 0.016\%   & DE 0.0628\%   & PL  0.045\%  & \\
CZ 0.005\%  & DE 0.016\%   & ES 0.0690\%   & AE 0.049\%   & \\
ID 0.005\%  & IE 0.042\%   & EU 0.478\%    & RS 0.049\%   & \\
NL 0.006\%  & SA  0.083\%  & US-CO 1.263\% & CA 0.059\%   & \\
PK 0.0139\% & GB 0.181\%   & US 3.542\%    & RU  0.11\%   & \\
CO 0.016\%  & US 0.276\%   & \textbf{GB 39.0897\%}  & FR 0.11\%    & \\
HK 0.0174\% & SG 0.340\%   & \textbf{FR 55.4101\%}  & US-WA 0.12\% & \\
RO 0.019\%  & ZA 0.574\%   &               & US  0.16\%   & \\
CR 0.02\%   & FR  1.167\%  &               & DE  0.17\%   & \\
BG 0.021\%  & KE 1.217\%   &               & MU  4.41\%   & \\
IT 0.021\%  & \textbf{MY 13.764\%}  &               & \textbf{GB  94.4\%}   & \\
VE 0.022\%  & \textbf{IT  32.228\%} &               &              & \\
NA 0.04\%   & \textbf{EU  50.004\%} &               &              & \\
SE 0.04\%   &              &               &              & \\
SG 0.042\%  &              &               &              & \\
RS 0.046\%  &              &               &              & \\
CA 0.059\%  &              &               &              & \\
TH 0.060\%  &              &               &              & \\
RU 0.096\%  &              &               &              & \\
ZA 0.125\%  &              &               &              & \\
DE 0.193\%  &              &               &              & \\
EU 0.250\%  &              &               &              & \\
US 0.602\%  &              &               &              & \\
US 4.39\%   &              &               &              & \\
\textbf{FR 10\%}    &              &               &              & \\
\textbf{GB 83.79\%} &              &               &              & \\
\hline
\end{tabular} 
\end{scriptsize}
\end{center}
\label{tab:exitpoint}
\end{table}

\begin{figure}[ht!]
    \begin{subfigure}[h]{.49\columnwidth}
     		\centering
		\includegraphics[height=0.98\textwidth, angle=-90]{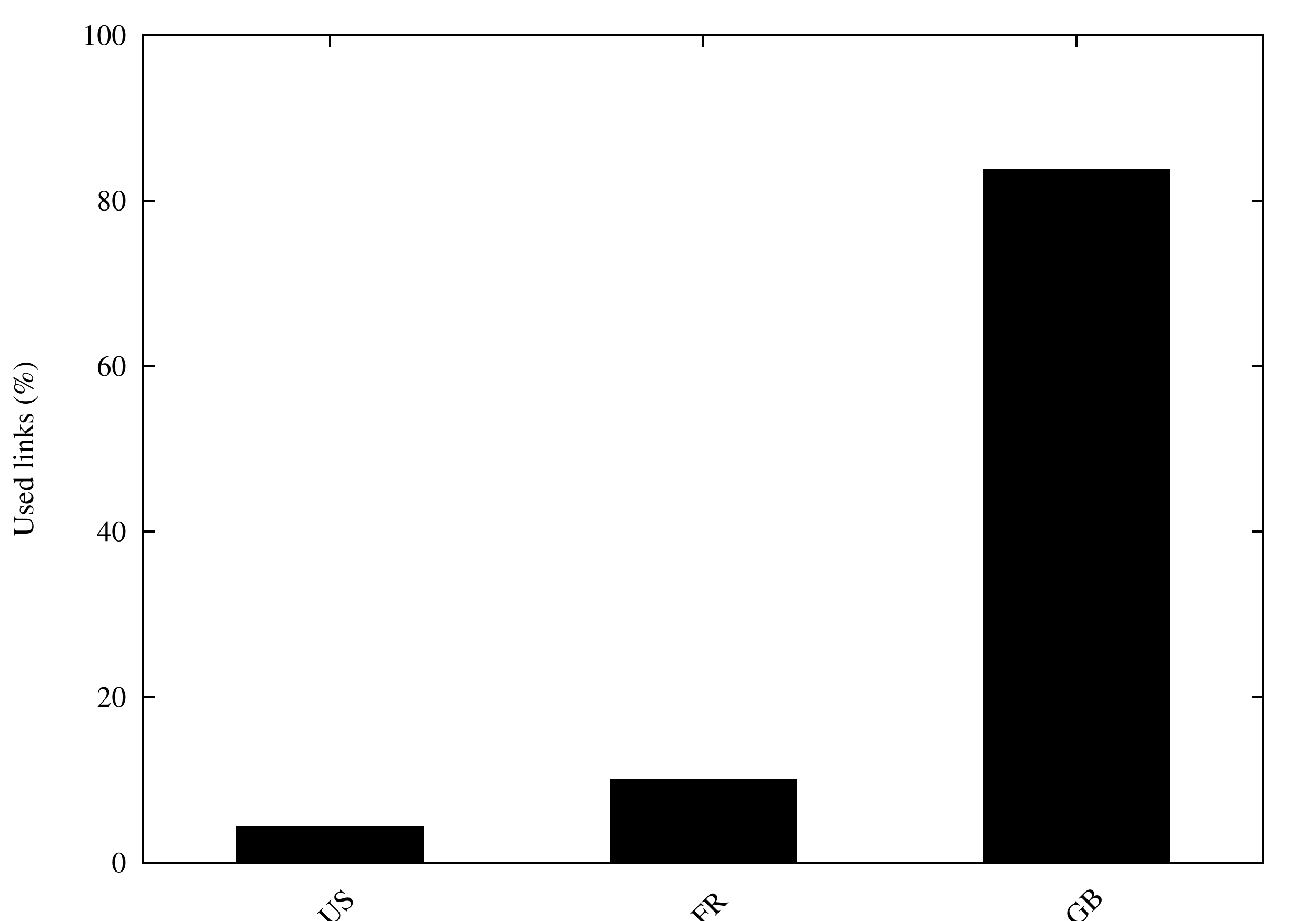}
		\caption{Sorted by countries}
		\label{fig:mg_pays}
    \end{subfigure}
    \begin{subfigure}[h]{.49\columnwidth}
    	\centering
		\includegraphics[height=0.98\textwidth, angle=-90]{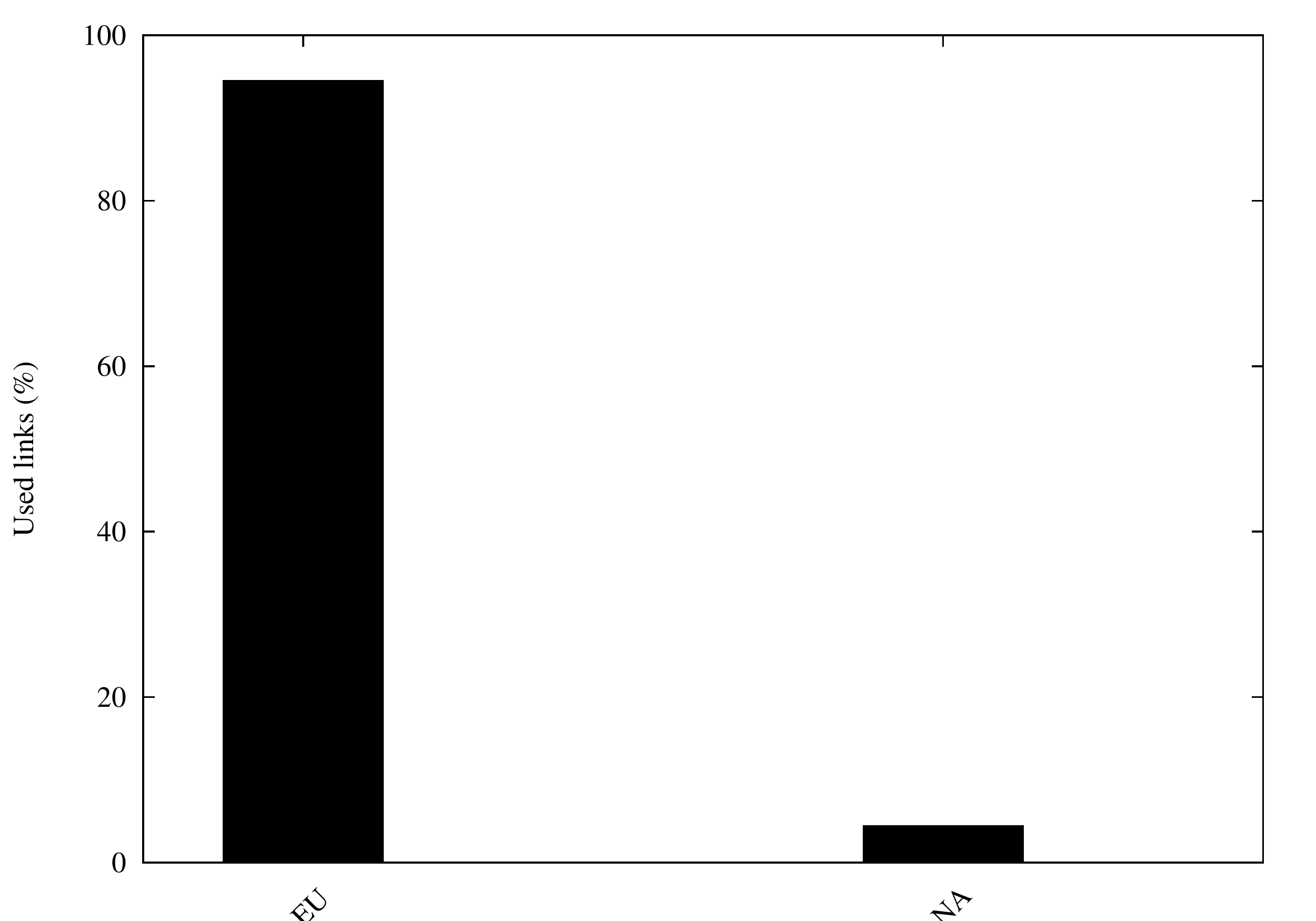}
		\caption{Sorted by continent}
		\label{fig:mg_continent}
    \end{subfigure}
    \caption{Number of Madagascar's Internet output links (up to 1\%)}
    \label{fig:mg}
\end{figure}

Figures \ref{fig:mg_pays} and~\ref{fig:mg_continent} give the repartition of exit points by country and by continent for Madagascar. In these figure we only plot the exit points that is used for at least 1\% of our data. We can see from these figures that only three countries are used as exit points for Madagascar and that more than 94\% of the traffic exists through Europe. When combined with the previous results, we can say that Europe has a well meshed network with all other countries since from IOA island point of view the entire world is at constant number of hop on average. These results also explain the decreasing behavior between the distance in kilometers and the number of hops. Since all traffic exists through Europe, reaching IOA island from Madagascar needs more hop count than reaching an European country.\\



\begin{figure}[ht!]
    \begin{subfigure}[h]{.49\columnwidth}
     		\centering
		\includegraphics[height=0.98\textwidth, angle=-90]{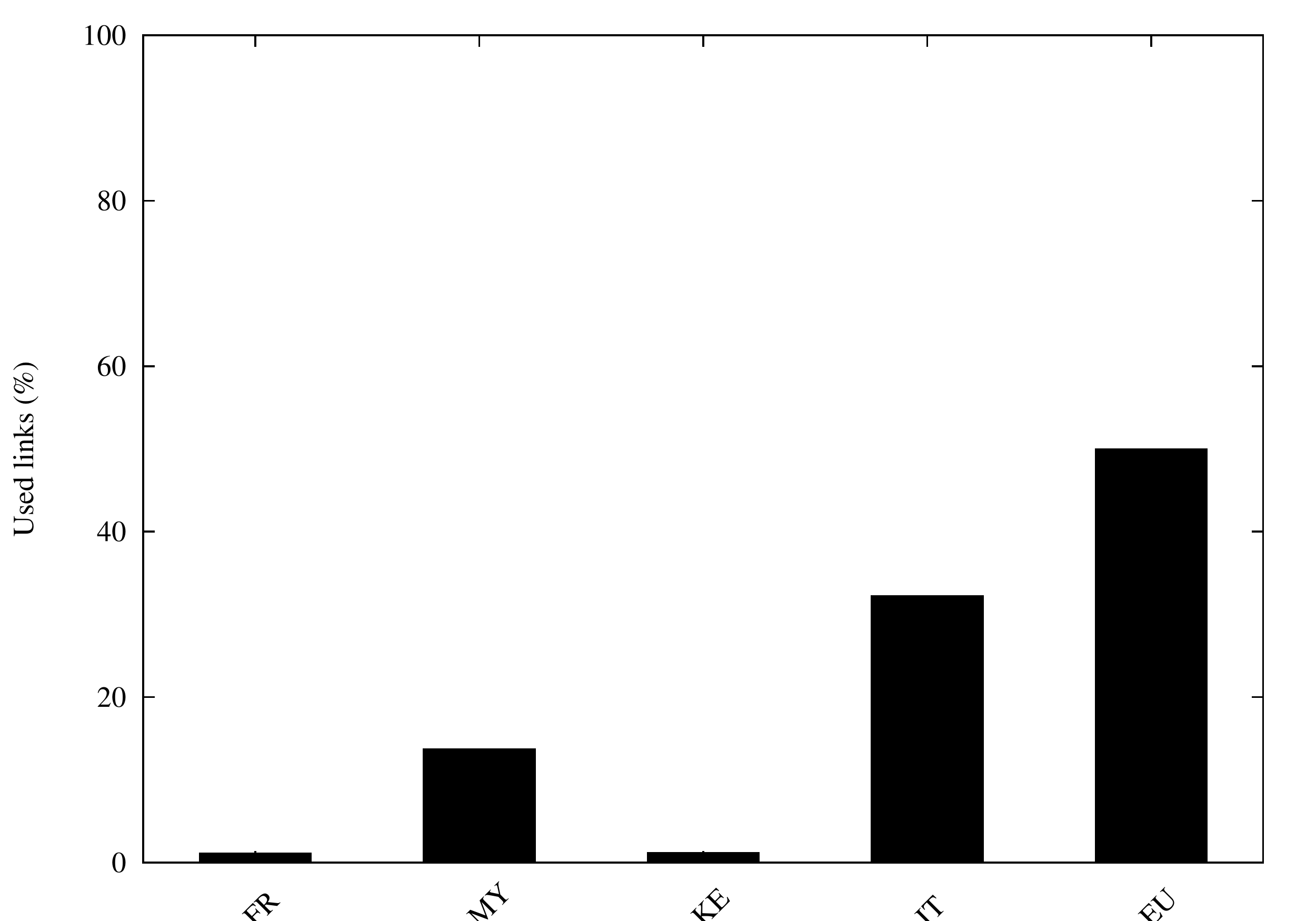}
		\caption{Sorted by countries}
		\label{fig:mu_pays}
    \end{subfigure}
    \begin{subfigure}[h]{.49\columnwidth}
    	\centering
		\includegraphics[height=0.98\textwidth, angle=-90]{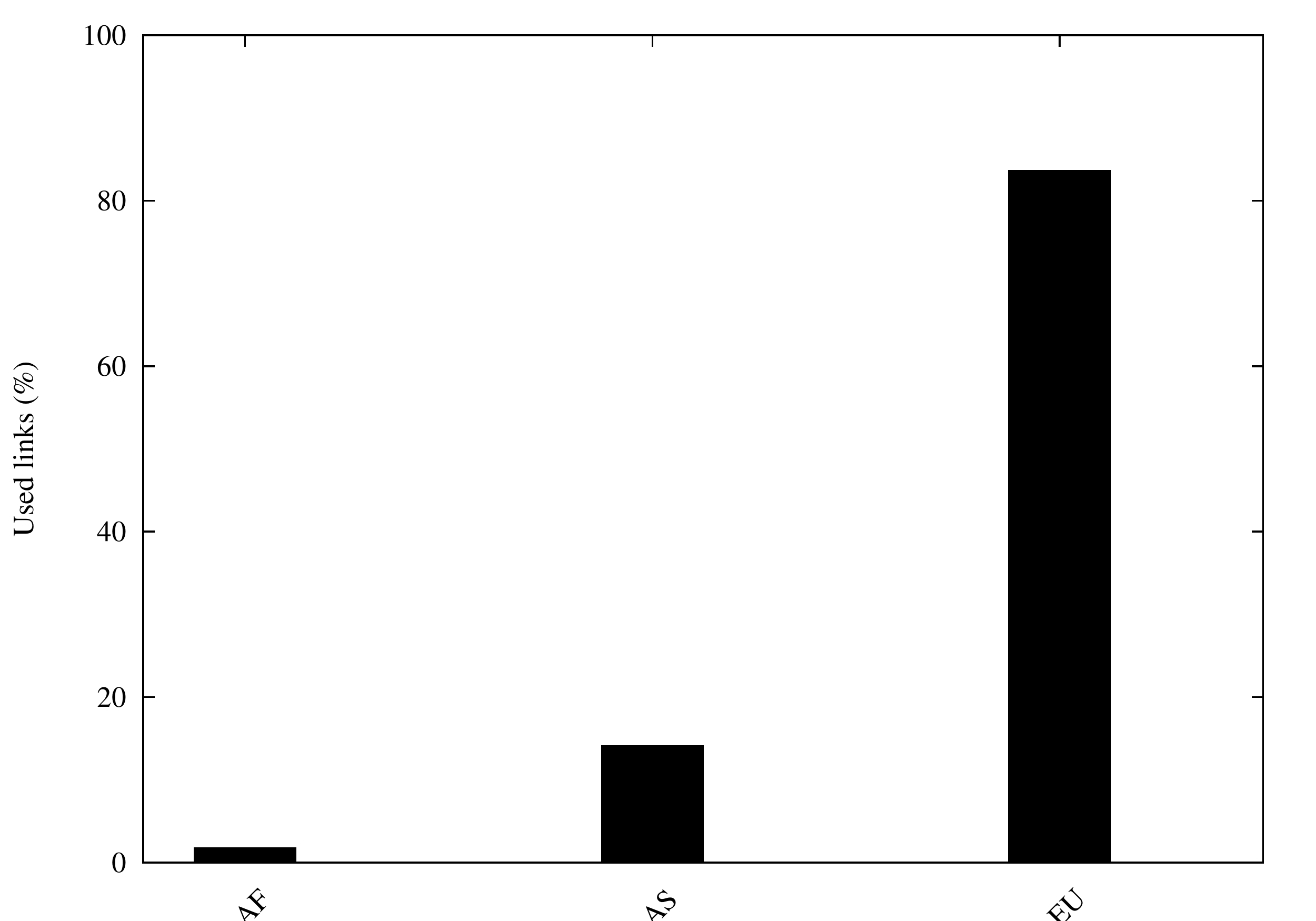}
		\caption{Sorted by continent}
		\label{fig:mu_continent}
    \end{subfigure}
    \caption{Number of Mauritius's Internet output links (up to 1\%)}
    \label{fig:mu}
\end{figure}

Figures~\ref{fig:mu_pays} and~\ref{fig:mu_continent} show the exit points for Mauritius. For Mauritius, the number of countries holding more that 1\% of the traffic is higher that Madagascar's. These figures show that more than 80\% of the traffic of Mauritius exit through Europe. However, it is is important to notice that unlike Madagascar, the most used exit point is not Great Britain.\\

\begin{figure}[ht!]
    \begin{subfigure}[h]{.49\columnwidth}
     		\centering
		\includegraphics[height=0.98\textwidth, angle=-90]{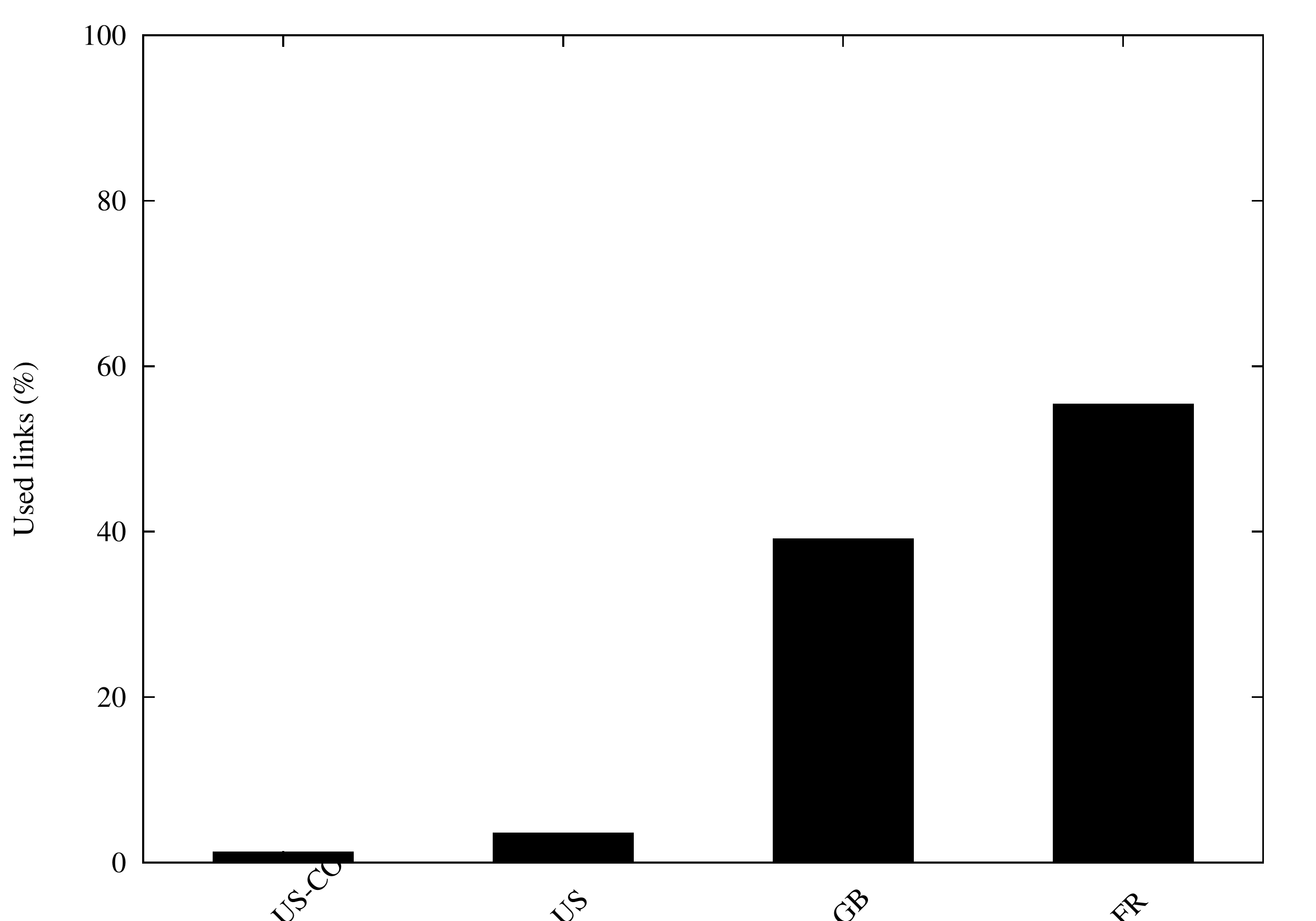}
		\caption{Sorted by countries}
		\label{fig:re_pays}
    \end{subfigure}
    \begin{subfigure}[h]{.49\columnwidth}
    	\centering
		\includegraphics[height=0.98\textwidth, angle=-90]{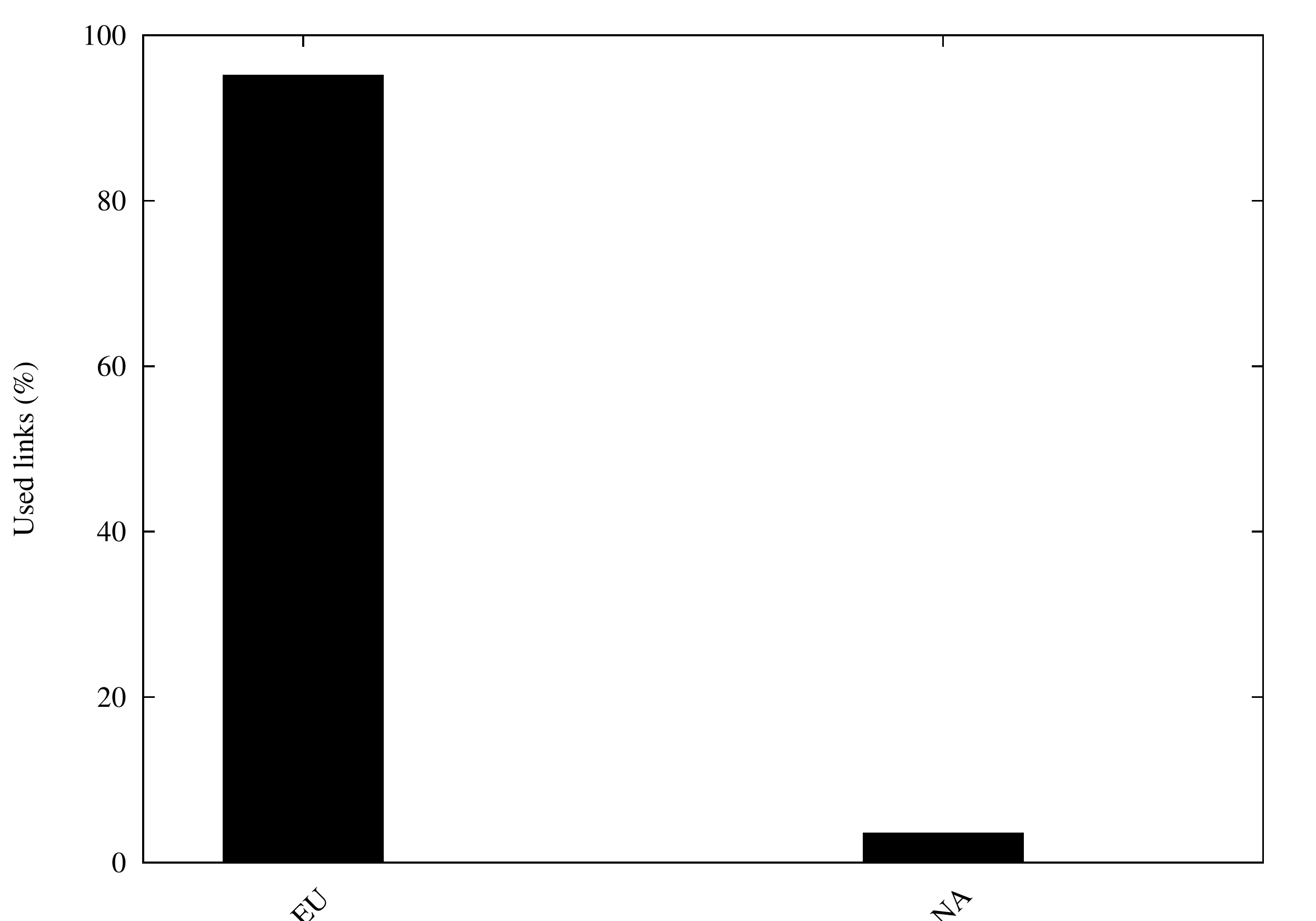}
		\caption{Sorted by continent}
		\label{fig:re_continent}
    \end{subfigure}
    \caption{Number of Reunion's Internet output links (up to 1\%)}
    \label{fig:re}
\end{figure}

Figures \ref{fig:re_pays} and~\ref{fig:re_continent} show the results of Reunion Island and we can see that the trend are the same as Madagascar and Mauritius trends. More than 95\% of the traffic exit through Europe and especially through France.\\

\begin{figure}[ht!]
    \begin{subfigure}[h]{.49\columnwidth}
     		\centering
		\includegraphics[height=0.98\textwidth, angle=-90]{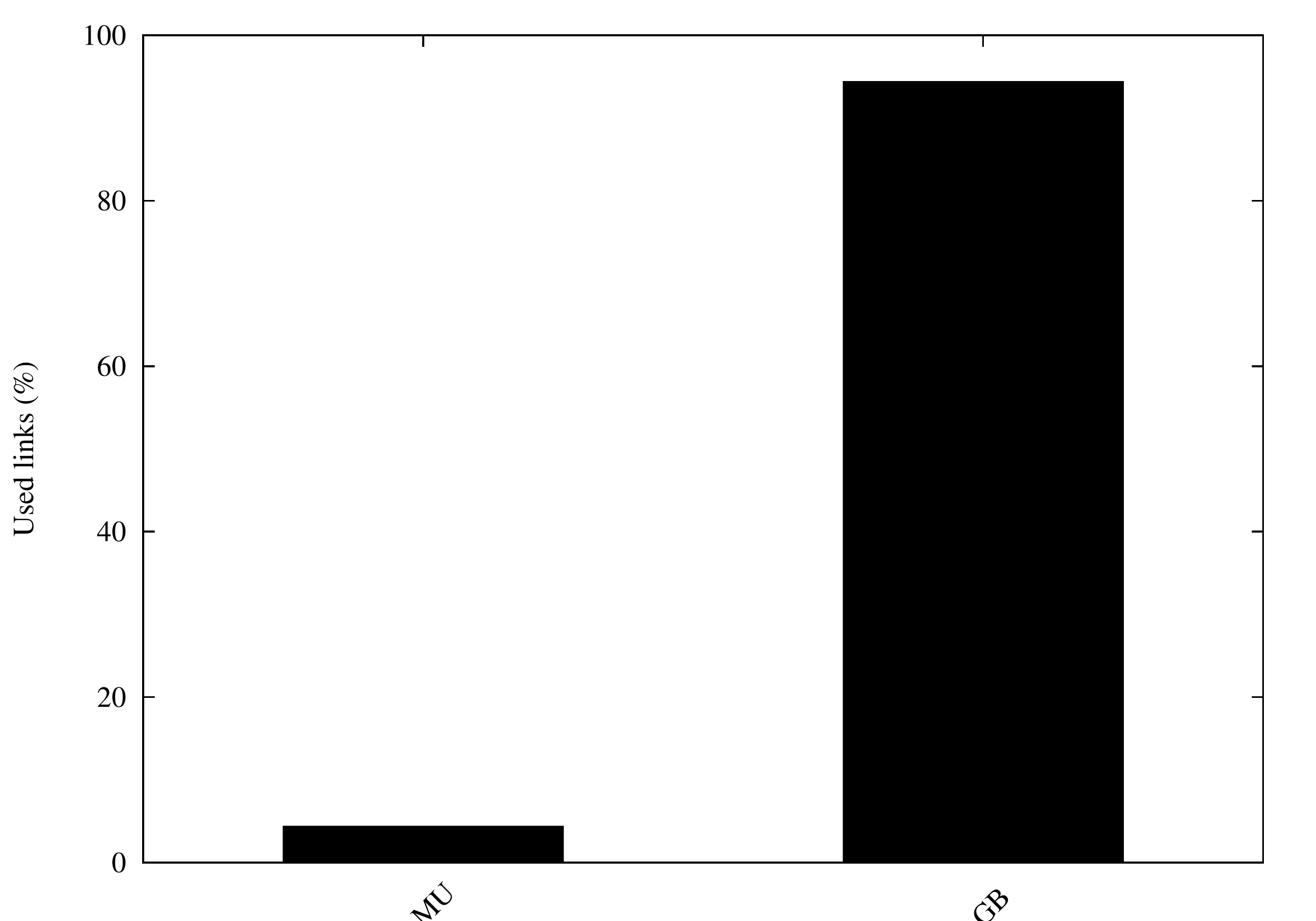}
		\caption{Sorted by countries}
		\label{fig:sc_pays}
    \end{subfigure}
    \begin{subfigure}[h]{.49\columnwidth}
    	\centering
		\includegraphics[height=0.98\textwidth, angle=-90]{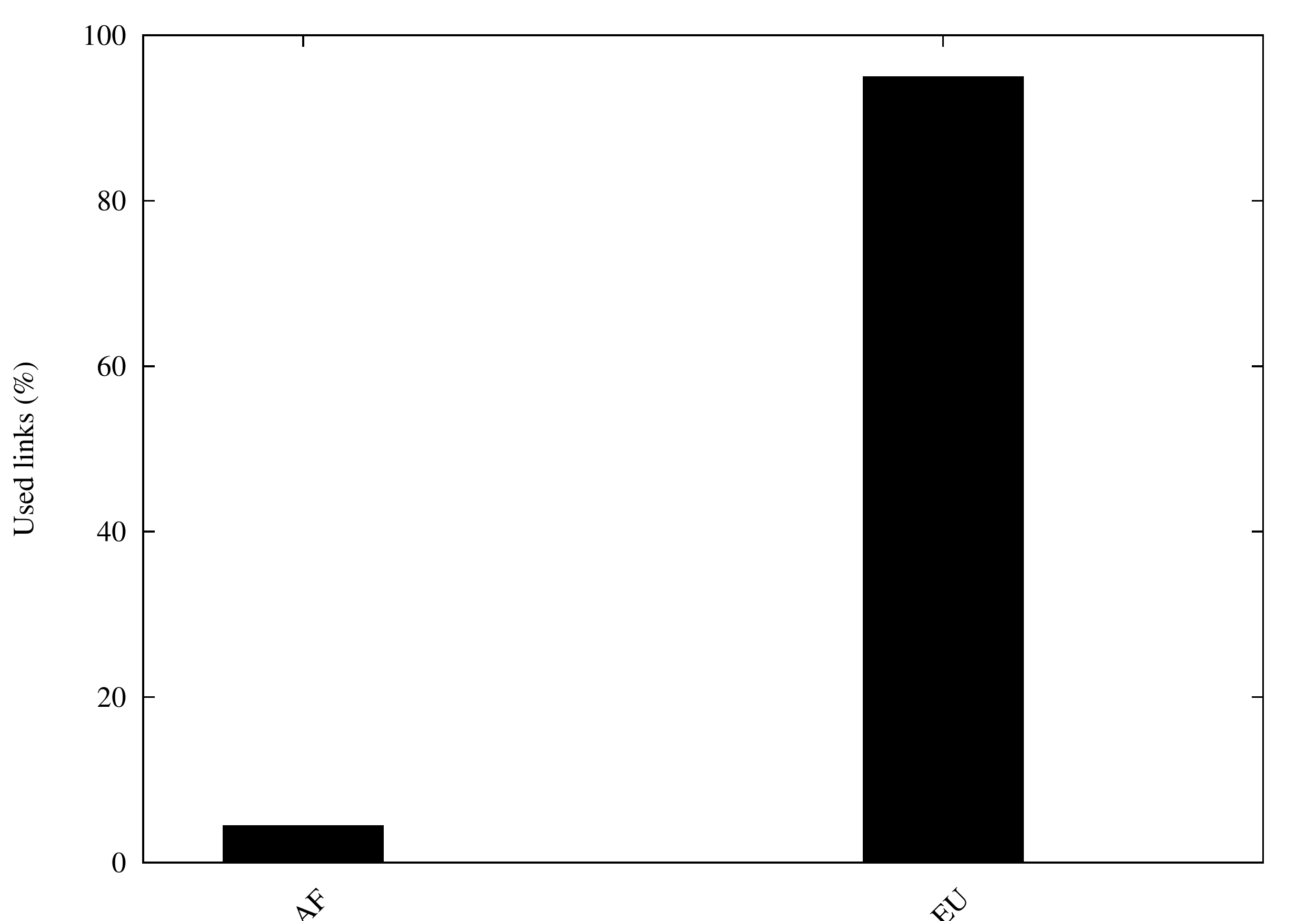}
		\caption{Sorted by continent}
		\label{fig:sc_continent}
    \end{subfigure}
    \caption{Number of Seychelles's Internet output links (up to 1\%)}
    \label{fig:sc}
\end{figure}

Figures \ref{fig:sc_pays} and~\ref{fig:sc_continent} show the results for Seychelles. We can see from these figures that more than 90\% of the traffic exists through Europe and especially Great Britain.\\

For Mayotte, none Figure are needed. All our data gone through Europe, and more specifically in France, as showned in Table \ref{tab:exitpoint}.We can see that Mayotte is a special case of IOA islands. Indeed, Mayotte uses only one exit point which is France. This make the Internet access of Mayotte not resilient/robust in case of failure. \\



In the previous results, showed in Figures \ref{fig:output} to \ref{fig:sc}, we can see that for traffic from IOA islands more than 80\% of the traffic use only one Continent exit point. Moreover, even if this exit point is in Europe for Madagascar, Reunion Island, Mauritius, Seychelles and Mayotte they are not all located in the same country which increase the RTT, number of hop and therefore reduce the Internet performance.


\section{Related work}
\label{sec:related}
The IOA routing rules increase delays. It is not the unique cases present in the World. 
Recent studies about routing rules show their impact on the delay. In \cite{zheng2005}, the authors work on the notion of {\em Triangle Inequality Violations} (TIV) and its impact on the delay. This notion said that the sum of delay between two nodes of a triangle is necessarily higher as the delay between one of the two previous nodes to the last one. If this rule is not respected, it is the case of TIV. One particular TIV is  called \emph{Boomerang routing}. A study of this phenomenon \cite{obar2012} as shown that many paths between Canadian ISP's take indirect paths through the USA. This sort of connection was frequent in IOA. Only the presence of the IXP and a real inter-connection of the ISP could resolve this problem.

For \cite{gupta2014} the main reason for long delay in the African region is due to peering agreements. Despite the numerous IXPs in South Africa or West Africa, some ISPs preferred to inter-connect in an European or Asian IXP. To bypass this rule, AFNIC and private companies, like Google, Akamai, etc... have made some investments in the African continent \cite{axis}. In \cite{fanou2015}, authors show that new infrastructures have not been correctly used by the African ISP. 

These needed to join an IXP based outside of the African continent, and that dependence on submarine cable. Chan and all worked on the impact of failures in submarine cables, in particularly on the SEA-ME-WE-4 \cite{Chan2011}. Furthermore adding a new submarine cable or increasing their bandwidth will not reduce latency \cite{RITE2014,telegeography}.

\section{Conclusion} 
\label{sec:conclusion}

Studying path and delay is a very important task in regions where the Internet access is not very fairly distributed. The Indian Ocean Area are connected to the Internet by only one or two submarines cables, depending on the country. From our probes, we used the \emph{paris-traceroute} tool to create an active metrology measurement, and used our tool \emph{rtraceroute} to analyze the data produced and enhance the knowledge between topology and logical paths.


Our results shows that the distance have no impact on the path length. For each new adding-node in the path, the increasing delay are variable, depended of the source's country. The surprising results is the decrease of the delay when the distance between the source and destination increases.  The major result indicates that most of the islands are world-connected but with a poor regional peering and meshing. It seems that the IXP and regional peering are not really optimized / well configured. We encourage the ISP for a better use of the IXP.
We discover that Mayotte is really a special case with only one direct connection to France. We are also surprised to discover that the only interconnection between the five islands \{MG, MU, SC, RE, YT\} is reduced to: \{$MG \rightarrow MU$\} and \{$SC \rightarrow MU$\}. It is a poor peering.

The future step of our research concern the deployment of probes in the other French overseas department and compare the situation. We leave for the future work the path and delay evolution over time in the IOA. An analysis of the TCP performance of the IOA could also be done, with the help of the different local ISP.
From this study, we can imagine to place a closest Regional IXP, to improve the regional peering and enhance the TCP performance.

\bibliography{../../../metrology}
\bibliographystyle{IEEEtran}

\end{document}